\documentstyle[sprocl]{article}

\def\Journal#1#2#3#4{{#1} {\bf #2}, #3 (#4)}

\def\NPB{{\em Nucl. Phys.} B}
\def\PLB{{\em Phys. Lett.}  B}
\def\PRL{\em Phys. Rev. Lett.}
\def\PRD{{\em Phys. Rev.} D}
\def\ZPC{{\em Z. Phys.} C}

\def\nc{N_c^{\rm eff}}

\def\la{\langle}
\def\ra{\rangle}
\def\non{\nonumber}

\def\be{\begin{equation}}
\def\ee{\end{equation}}
\def\bea{\begin{eqnarray}}
\def\eea{\end{eqnarray}}

\begin{document}

\title{Exclusive charmless $B_s$ hadronic decays into $\eta'$ and $\eta$}

\author{B. Tseng}

\address{Institute of Physics, Academia Sinica, Taipei, Taiwan 115, R.O.C.
\\E-mail: btseng@phys.sinica.edu.tw}


\maketitle\abstracts{ Using the next-to-leading order
QCD-corrected effective Hamiltonian, charmless exclusive
nonleptonic decays of the $B_s$ meson  into $\eta$ or $\eta'$ are
calculated within the generalized factorization approach.
Nonfactorizable contributions are included with two different
treatments. Some subtleties involved are  discussed.}

\section{Introduction}
Stimulated by the recent observations of the large inclusive and
exclusive rare $B$ decays by the CLEO Collaboration \cite{CLEO},
there are considerable interests in the charmless $ B $ meson
decays \cite{Chau1}. To explain the abnormally large  branching
ratio of the semi-inclusive process $B\to\eta'+X$, several
mechanisms have been advocated \cite{AS,HT,Frit,mechanism}  and
some tests of these mechanisms have been proposed \cite{DuYang97}.
It is now generally believed that the QCD anomaly
\cite{AS,HT,Frit} plays a vital role. The understanding of the
exclusive $B \to \eta' K$, however, relies on several subtle
points. First, the QCD anomaly does occur through the equation of
motion \cite{KP97,Ali97} when  calculating the $(S-P)(S+P)$
penguin operator and its effect is found to reduce the branching
ratio. Second, the mechanism of $c\bar{c} \to\eta'$, although
proposed to be large and positive originally \cite{HZ97,SZ98}, is
now preferred to be negative and smaller than before as implied by
a recent theoretical recalculation \cite{AMT} and several
phenomenological analyses \cite{Ali97,FK97}. Third, the running
strange quark mass which appears in the calculation of the matrix
elements of the $(S-P)(S+P)$ penguin operator, the $SU(3)$
breaking effect in the involved $\eta'$ decay constants and the
normalization of the $B \to \eta^{(')}$ matrix element involved
raise the branching ratio substantially. Finally, nonfactorizable
contributions, which are parametrized by the $N_c^{\rm eff}$,
gives the final answer for the largeness of exclusive $B \to \eta'
K$~\cite{CT98,ALI98}. It is very interesting to see the impacts of
these subtleties mentioned above on the the exclusive charmless
$B_s$ decays to an $\eta'$ or $\eta$~\cite{BS}. That is the main
purpose of this talk~\cite{BS1998}.

\section{\it Theoretical Framework}
We begin with a brief description of the theoretical framework.
The relevant effective $\Delta B=1$ weak Hamiltonian is
\be
{\cal H}_{\rm eff} = {G_F\over\sqrt{2}}\Big[ V_{ub}V_{uq}^*(c_1
O_1^u+c_2O_2^u)+V_{cb}V_{cq}^*(c_1O_1^c+c_2O_2^c)
-V_{tb}V_{tq}^*\sum^{10}_{i=3}c_iO_i\Big], \ee
 where
$q=d,s$, and $O_{3-6}$ are QCD penguin operators and $O_{7-10}$
are electroweak penguin operators. $C_i(\mu)$ are the Wilson
coefficients, which have been evaluated to the next-to-leading
order (NLO) \cite{Buras92,Ciuchini}. One important feature of the
NLO calculation is the renormalization-scheme and -scale
dependence of the Wilson coefficients (for a review, see
\cite{Buras96}). In order to ensure the $\mu$ and renormalization
scheme independence for the physical amplitude, the matrix
elements, which are evaluated under the factorization hypothesis,
have to be computed in the same renormalization scheme and
renormalized at the same scale as $c_i(\mu)$. However, as
emphasized in \cite{CT98},  the matrix element $\la O\ra_{\rm
fact}$ is scale independent under the factorization approach and
hence it cannot be identified with $\la O(\mu)\ra$. Incorporating
QCD and electroweak corrections to the four-quark operators, we
can redefine $c_i(\mu)\la O_i(\mu)\ra= {c}_i^{\rm eff}\la
O_i\ra_{\rm tree}$, so that ${c}_i^{\rm eff}$ are  renormaliztion
scheme and scale independent. Then the factorization approximation
is applied to the hadronic matrix elements of the operator $O$ at
the tree level. The numerical values for ${c}_i^{\rm eff}$ are
shown in  the last column of Table I, where $\mu={m}_b(m_b)$,
$\Lambda^{(5)}_{\overline{\rm MS}}=225$ MeV, $m_t=170$ GeV and
$k^2=m_b^2/2$ are used \cite{CT98}.


In  general, there are contributions from the nonfactorizable
amplitudes. Because there is only one single form factor (or
Lorentz scalar) involved in the decay amplitude of $B\,(D)\to
PP,~PV$ decays ($P$: pseudoscalar meson, $V$: vector meson), the
effects of nonfactorization can be lumped into the effective
parameters $a_i^{\rm eff}$ \cite{Cheng}:
\be
a_{2i}^{\rm eff}=c_{2i}^{\rm eff}+c_{2i-1}^{\rm eff}\left({1\over
N_c}+\chi_{2i} \right),\qquad a_{2i-1}^{\rm eff}=c_{2i-1}^{\rm
eff}+c_{2i}^{\rm eff}\left({1\over N_c}+\chi_{2i-1}\right), \ee
where $c_{2i,2i-1}^{\rm eff}$ are the Wilson coefficients of the
4-quark operators, and nonfactorizable contributions are
characterized by the parameters $\chi_{2i}$ and $\chi_{2i-1}$. We
can parametrize the nonfactorizable contributions by defining an
effective number of colors $N_c^{\rm eff}$, called $1/\xi$ in
\cite{BSW}, as $ 1/N_c^{\rm eff} \equiv (1/N_c)+\chi$. Different
factorization approach used in the literature can be classified by
the effective number of colors $N_c^{\rm eff}$. The so-called
``naive" factorization discards all the nonfactorizable
contributions and takes $ 1/N_c^{\rm eff}= 1/N_c=1/3 $, whereas
the ``large-$N_c$ improved" factorization \cite{Buras} drops out
all the subleading $1/N_c$ terms and takes $ 1/N_c^{\rm eff}=0$.
In principle, $N_c^{\rm eff}$ can vary from channel to channel, as
in the case of charm decay. However, in the energetic two-body $B$
decays, $\nc$ is expected to be process insensitive as supported
by data \cite{Neubert}. If $N_c^{\rm eff}$ is process independent,
then we have a generalized factorization. In this paper, we will
treat the nonfactorizable contributions with two different
phenomenological ways
: (i) the one with ``homogenous" structure, which
assumes that  $(N_c^{\rm eff})_1  \approx (N_c^{\rm eff})_2
\approx \cdots \approx (N_c^{\rm eff})_{10}$ , and (ii) the
``heterogeneous" one, which considers the possibility of
$N_c^{\rm
eff}(V+A)\neq N_c^{\rm eff}(V-A)$.
 The consideration of the
``homogenous" nonfactorizable contributions, which is commonly
used in the literature, has its advantage of simplicity. However,
as argued in \cite{CT98}, due to the different Dirac structure of
the Fierz transformation, nonfactorizable effects in the matrix
elements of $(V-A)(V+A)$ operators are {\it a priori} different
from that of $(V-A)(V-A)$ operators, i.e. $\chi(V+A)\neq
\chi(V-A)$. Since $1/N_c^{\rm eff}=1/N_c+\chi$ , theoretically it
is expected that
\bea
 N_c^{\rm eff}(V-A)&\equiv& \left(N_c^{\rm
eff}\right)_1\approx\left(N_c^{\rm eff}\right)_2\approx
\left(N_c^{\rm eff}\right)_3\approx\left(N_c^{\rm
eff}\right)_4\approx \left(N_c^{\rm eff}\right)_9\approx
\left(N_c^{\rm eff}\right)_{10},   \non \\
 N_c^{\rm eff}(V+A)&\equiv& \left(N_c^{\rm eff}\right)_5\approx\left(N_c^{\rm
eff}\right)_6\approx \left(N_c^{\rm eff}\right)_7\approx
\left(N_c^{\rm eff}\right)_8, \eea
 To illustrate the effect of the
nonfactorizable contribution, we extrapolate  $N_c(V-A) \approx 2$
from $B \to D \pi(\rho)$ \cite{CT95} to charmless decays.

\begin{table}[ht]
\begin{center}
\caption{ Numerical values of effective coefficients $a_i$ at
$N_c^{\rm eff}=2,3,5,\infty$, where $N_c^{\rm eff}=\infty$
corresponds to $a_i^{\rm eff}=c_i^{\rm eff}$. The entries for
$a_3$,...,$a_{10}$ have to be multiplied with
$10^{-4}$.\label{tab:effa}}
\begin{tabular}{|c|c|c|c|c|}
\hline
 & $N_c^{\rm eff}=2$ & $N_c^{\rm eff}=3$ & $N_c^{\rm eff}=5$  &
$N_c^{\rm eff}=\infty$ \\ \hline $a_1$  &0.986 & 1.04  & 1.08 &
1.15  \\ $a_2$  &0.25 &  0.058 & -0.095 &  -0.325 \\ $a_3$
&$-13.9-22.6i$   & 61 & $120+18i$ &$211+45.3i$ \\ $a_4$
&$-344-113i$ & $-380-120i$  & $-410-127i$ & $-450-136i$ \\ $a_5$
&$-146-22.6i$ & $-52.7$  & $22+18i$ & $134+ 45.3i$ \\ $a_6$  &$-
493-113i$ & $-515-121i$  & $-530-127i$ & $-560-136i$ \\ $a_7$  &$
0.04-2.73i$ & $-0.7-2.73i$  & $-1.24-2.73i$ & $-2.04-2.73i$  \\
$a_8$  &$2.98 -1.37i$ & $3.32-0.9i$  & $3.59-0.55i$ & 4   \\ $a_9$
&$-87.9- 2.73i$ & $-91.1-2.73i$  & $-93.7-2.73i$ & $-97.6-2.73i$
\\ $a_{10}$&$-29.3-1.37i$ & $-13-0.91i$  & $-0.04-0.55i$ & 19.48
\\ \hline
\end{tabular}
\end{center}
\end{table}
 The
$N_c^{\rm eff}$-dependence of the effective parameters $a_i$'s are
shown in Table I, from which we see that $a_1, a_4, a_6$ and $a_9$
are $N_c^{\rm eff}$-stable, and the remaining ones are $N_c^{\rm
eff}$-sensitive. We would like to remark that while $a_3$ and
$a_5$ are both $N_c^{\rm eff}$-sensitive, the combination of ($a_3
- a_5$) is rather stable under the variation of the $N_c^{\rm
eff}$ within the ``homogeneous" picture and is still sensitive to
the factorization approach taken in the ``heterogeneous" scheme.
This is the main difference between the ``homogeneous"  and
``heterogeneous" approaches. While $a_7,a_8$ can be neglected,
$a_3,a_5$ and $a_{10}$ have some effects on the relevant processes
depending on the choice of $N_c^{\rm eff}$.

\section{ Phenomenology}

\begin{table}[h]
\caption{Average branching ratios (in units of $10^{-6}$) for
charmless $B_s$ decays to $\eta'$ and $\eta$. Predictions are for
$k^2=m_b^2/2$, $\eta=0.34,~\rho=0.16$. I denotes the
``homogeneous" nonfactorizable contributions {\rm i.e. $N_c^{\rm
eff}(V-A)=N_c^{\rm eff}(V+A)$} and (a,b,c,d) represent the cases
for $N_c^{\rm eff}$=($\infty$,5,3,2). II denotes the
``heterogeneous" nonfactorizable contributions, {\rm i.e.
$N_c^{\rm eff}(V-A) \neq N_c^{\rm eff}(V+A)$ } and
($a'$,$b'$,$c'$) represent the cases for $N_c^{\rm
eff}(V+A)$=(3,5,$\infty$), where we have fixed $N_c^{\rm
eff}(V-A)$=2 (see the text) \label{tab:br}}

\footnotesize
\begin{center}
\begin{tabular}{|l|c c c c |c c c |}
\hline Decay & $I_a$ & $I_b$ & $I_c$ & $I_d$ & $II_{a'}$ &
$II_{b'}$ & $II_{c'}$ \\ \hline $ \bar{B_s}\to\pi \eta' $ & 0.25 &
0.17 &0.13&0.11 &0.11&0.11&0.10 \\ $ \bar{B_s}\to\pi \eta $  &
0.16 & 0.11& 0.08 & 0.07 & 0.07&0.068&0.067 \\ $ \bar{B_s}\to \rho
\eta' $ & 0.70&0.47&0.36&0.30&0.30&0.30&0.31 \\ $ \bar{B_s}\to\rho
\eta $ & 0.45&0.30&0.24&0.19&0.19&0.19&0.20 \\ $ \bar{B_s}\to
\omega \eta' $ & 6.9&0.9&0.012&2.14&0.48&0.03&0.83 \\ $
\bar{B_s}\to \omega \eta $ &4.45 &0.63&0.008&1.39&0.31&0.02&0.54
\\ $ \bar{B_s}\to\eta' K^0$ &1.25&1.07&1.01&1.00&1.27&1.51&1.90 \\
$ \bar{B_s}\to\eta K^0$ &1.35 &0.81&0.68&0.76&0.75 &0.74&0.72\\ $
\bar{B_s}\to \eta' K^{*0}$ & 0.49 & 0.35 &0.32 &0.26 &
0.49&0.60&0.80\\ $ \bar{B_s}\to\eta K^{*0}$ &0.45 & 0.05  & 0.02 &
0.24 & 0.24&0.24&0.25\\ $ \bar{B_s}\to\eta \eta'$ & 47.4&41.8
&38.3 &34.4 &39.5&44.1&51.5 \\ $ \bar{B_s}\to\eta' \eta'$ & 26.6 &
24.9&23.8 &22.4&33.8&43.9&62.2 \\ $ \bar{B_s}\to\eta \eta$ &20.3
&17.1 &15.1 &12.8&11.6&10.7&9.1 \\ $ \bar{B_s}\to\phi \eta'$ &
0.44 &0.59 &2.29 &6.20&4.41&3.11&1.66 \\ $ \bar{B_s}\to\phi \eta$
& 0.04 &0.91 &2.29 &4.92&2.28 &0.92&0.10   \\ \hline
\end{tabular}
\end{center}
\end{table}
With the following input parameters, we obtain the branching
ratios shown in Table~\ref{tab:br}.
\begin{itemize}
\item For the running quark masses, we use \cite{Fusaoku}
\bea
 m_u(m_b)&=&3.2\,{\rm MeV},  ~~m_d(m_b)=6.4\,{\rm MeV}, \qquad
m_s(m_b)=105\,{\rm MeV},  \non \\
  m_c(m_b)&=&0.95\,{\rm GeV},
~~ m_b(m_b)=4.34\,{\rm GeV}, \eea
\item The Wolfenstein parameters with $A=0.81$ ,
$\lambda=0.22$, $\rho=0.16$, and $\eta=0.34$ are used in this
work.
\item For values of the decay constants, we use $f_\pi=132$ MeV, $f_ K=160$
MeV, $f_\rho=210$ MeV,  $f_{K^*}=221$ MeV,  $f_ \omega=195$ MeV
and $f_\phi=237$ MeV. For the matrix element, we use the
relativistic quark model's results \cite{CCW} with a proper
normalization.
\end{itemize}

From this studies, we learned that
\begin{itemize}
\item Similar to their $B_{u,d}$ corresponding decay modes, $B_s
\to \eta{(')} \eta{(')}$ have the largest branching ratios
($O(10^{-5}$) and thus are the interesting modes to be observed in
the near future.
\item Since the internal W-emission is CKM-suppressed and the QCD penguins are
canceled  out in these decay modes,
$\bar{B_s}\to\pi(\rho)\eta^{(')}$ are dominated by the EW penguin
diagram. The dominant EW penguin contribution proportional to
$a_9$ is $N_c^{\rm eff}$-stable. Thus, by measuring these
branching ratios, we can determine the effective coefficient
$a_9$.
\item It is found that for processes
depending on the $N_c^{\rm eff}$-stable $a_i$'s such as $\bar{B_s}
\to (\pi , \rho) \eta^{(')}$, the branching ratios are not
sensitive to the factorization approach we used. While for the
processes depending on  the $N_c^{\rm eff}$-sensitive $a_i$'s such
as the $\bar{B_s} \to \omega \eta^{(')}$, the predicted branching
ratios have a wide range depending on the choice of the
factorization approach. It means that even within the standard
model, there are large uncertainties for these $N_c^{\rm
eff}$-sensitive processes.
\item   For the mechanism $(c\bar c) \to \eta'$ ,in general, it has
smaller effects  due to a possible CKM-suppression and the
suppression in the  decay constants except for the $\bar{B_s} \to
\phi \eta$ under the ``large-$N_c$ improved" factorization
approach, where the internal $W$ diagram is CKM-suppressed and the
penguin contributions are compensated.
\end{itemize}

\section{\it Summary and Discussions}
We have studied charmless exclusive nonleptonic $B_s$ meson decay
into an $\eta$ or $\eta'$  within the generalized factorization
approach. Nonfactorizable contributions are parametrized in terms
of  the effective number of colors $N_c^{\rm eff}$ and predictions
using different factorization approaches are shown with the
$N_c^{\rm eff}$ dependence.

In our work, we, following the standard approach, have neglected
the $W$-exchange and the space-like penguin contributions. Another
major source of uncertainties comes from the form factors we used,
which are larger than the BSW model's calculations. For simple
processes such as $B_s \to \pi(\rho,\omega) \eta{(')}$, they only
scale with a factor, while for the complicated processes like $B_s
\to K^0 \eta{(')}$ the different contributions (tree, QCD penguin,
EW penguin) will have different weights. Although the Wolfenstein
parameter $\rho$ ranges from the negative region to the positive
one, we have ``fixed" it to some representative values. The
interference pattern between the internal $W$ diagram and the
penguin contributions will change when we take a different sign of
$\rho$.

\section*{Acknowledgments}
I  would like to thank  Prof. Guey-Lin Lin for  this stimulating
workshop. This work is supported in part by the National Science
Council of the Republic of China under Grant NSC87-2112-M006-018.

\section*{References}
\newcommand{\bi}{\bibitem}

\end{document}